\documentclass{rsproca_new}
\usepackage{braket}

\usepackage{breqn}
\usepackage{comment}

\begin{document}

\title{Magic State Distillation with the Ternary Golay Code}
\author{Shiroman Prakash}

\address{Department of Physics and Computer Science,\\Dayalbagh Educational Institute,\\Dayalbagh, Agra, India 282005}
\corres{Shiroman Prakash\\
\email{sprakash@dei.ac.in}}

\begin{abstract}
The ternary Golay code -- one of the first and most beautiful classical error-correcting codes discovered  -- naturally gives rise to an 11-qutrit quantum error correcting code. We apply this code to magic state distillation, a leading approach to fault-tolerant quantum computing. We find that the 11-qutrit Golay code can distill the ``most magic'' qutrit state -- an eigenstate of the qutrit Fourier transform known as the \textit{strange state} -- with cubic error-suppression and a remarkably high threshold. It also distills the ``second-most magic'' qutrit state, the Norell state, with quadratic error-suppression and an equally high threshold to depolarizing noise. 
\end{abstract}

\maketitle

\section{Introduction} 

The classical Golay codes \cite{golay, golayUnique} are amongst the first and most beautiful ways discovered to protect classical information. Two Golay codes exist -- the 23-bit binary Golay code and the 11-trit ternary Golay code. These codes are unique, in that they are the only linear perfect classical error correcting codes other than the Hamming codes. While they were discovered through a computer search, (and independently by a Finnish football enthusiast, apparently via trial and error), their discovery led to profound advancements in the theory of coding as well as the mathematical theory of finite groups.\cite{Barg}

Can the Golay codes provide us better ways to protect quantum information from noise? Via the CSS construction, the Golay codes can be used to construct  $[23,1,7]_2$ and $[11,1,5]_3$  quantum error correcting codes. Applications of the 23-qubit Golay code to fault-tolerant quantum computing exist \cite{PhysRevA.68.042322,PaetznickGolay}, but the 11-qutrit Golay code has apparently never been studied. Here, we observe that the 11-qutrit Golay code is remarkably well-suited for a promising approach to fault-tolerant quantum computing known as magic state distillation\cite{MSD, knill}.

Magic state distillation \cite{MSD, knill, Reichardt2005, reichardt2009quantum} is a leading approach to fault tolerant quantum computing. In the past few years, magic state distillation for qudits of (typically odd prime) dimensions other than two has attracted some interest \cite{ACB, CampbellAnwarBrowne, campbell2014enhanced, Howard, qudit-bound-states}, and notably has been used to identify contextuality as an essential resource for universal quantum computation \cite{nature}. However, for the most part, qudit fault-tolerant quantum computing \cite{Gottesman1999} appears relatively unexplored, although attractive experimental realizations of qutrits do exist, e.g., \cite{PhysRevA.67.062313, PhysRevLett.105.223601, mair2001entanglement}.

In the magic state model, a fault-tolerant quantum computer has the ability to measure and initialize states without error in the computational basis, and act without error on these states with a discrete subgroup of the full set of unitary operators known as the Clifford group \cite{PhysRevA.57.127,Gottesman1999}. A quantum computer with only these capabilities is  classically simulable \cite{gottesman1998heisenberg, PhysRevA.70.052328, Veitch_2012, PhysRevLett.109.230503} and therefore not sufficient for universal quantum computation. In addition, the computer is able to prepare ancilla qudits in certain non-stabilizer states, called \textit{magic states}; but these states are produced with limited fidelity. To approximate a universal quantum computer within this model, we require arbitrarily pure magic states, which can be used to implement non-Clifford gates via state-injection. Using many low-fidelity magic states, it is sometimes possible to distill a small number of high-fidelity magic states via protocols involving only Clifford unitaries and stabilizer measurements. This process is only successful if the noise level of the low-fidelity input qudits is below a particular threshold associated with the particular distillation protocol employed. An open problem is to design a distillation protocol with as high a threshold as possible.

What constitutes a magic state for a qutrit? In entanglement theory, any state that is not a separable state is defined to be entangled. By analogy, any (pure) state that is not a stabilizer state is defined to be magic \cite{Veitch_2014}.  One can then ask, which qutrit state is \text{most} magic? To answer this question, a natural measure to use is the \textit{regularized entropy of magic}, which is defined as the relative entropy between a large supply of qutrits in the candidate magic state and the nearest multi-qutrit stabilizer state. Unfortunately, the regularized entropy of magic is not feasible to compute. To place rigorous bounds on magic, two useful surrogate measures exist: the \textit{mana},\cite{Veitch_2014} which is essentially a measure of the sum of negative entries in the discrete Wigner function \cite{Wootters1987, PhysRevA.70.062101, Appleby1, cormick2006classicality, Gross} of the candidate magic state; and the (regularized) \textit{thauma} \cite{wang2018efficiently} which is the minimum relative entropy between the many copies of the candidate magic state and a subnormalized state with positive Wigner function. 

Two qutrit magic states were identified in \cite{Veitch_2014} that maximize the mana -- the strange state, an eigenstate of the qutrit Fourier transform (which was first discovered in \cite{noiseQudit}), and the so-called Norell state, which is the eigenstate of another single qutrit Clifford operator $N$ defined below.  It was recently shown that the strange state has larger thauma \cite{wang2018efficiently} than the Norell state, hence earning it the accolade of the ``most magic'' qutrit state. This accolade is conceptually satisfying because the strange state also maximally violates the contextuality inequality of \cite{nature}, and is also the qutrit state for which distillation could be most robust to depolarizing noise. As we show in \cite{jain}, the qutrit strange state is also the most symmetric of all qudit magic states, and has no natural analogue in higher odd-prime dimensions.

Distillation of the strange state is an exciting problem for both practical and theoretical reasons. The strange state is furthest from the Wigner polytope \cite{Veitch_2012}, and therefore has potential to be distilled with the greatest threshold to noise of any qutrit state, as first observed in \cite{noiseQudit}. Moreover, constructing a magic state distillation routine that distills the strange state, with a threshold meeting the theoretical upper bound set by negativity of the Wigner function, would be tantamount to a proof that contextuality is sufficient for universal quantum computation, by the results of \cite{nature}. There are indications that this may be an impossible problem to solve -- it has be shown in \cite{qudit-bound-states} that no distillation protocol based on a stabilizer code of finite length can meet this upper bound. But \cite{qudit-bound-states} does not rule out the existence of an infinite sequence of protocols based on stabilizer codes of increasing length, whose threshold approaches this upper bound, in the asymptotic limit.

Previous works on qutrit and qudit magic state distillation \cite{CampbellAnwarBrowne, campbell2014enhanced, HowardVala} have mostly focused on distilling a class of equatorial magic states, which posses several useful properties 
\cite{maxnonlocality}, although they have non-maximal mana. In addition, eigenstates of the qutrit Fourier transform other than the strange state were distilled via the 5-qutrit code in \cite{ACB}, and \cite{Howard} presented a distillation routine for the qutrit Norell state; although one should note that the protocols of \cite{ACB}, and \cite{Howard} have only a linear reduction in noise rate. Prior to this work, no magic state distillation routine with the strange state as a stable endpoint was known. 

Here, we show that an $[11,1,5]_3$  code obtained from the ternary Golay code distills both the Norell state and the strange state, with a threshold to depolarizing noise that exceeds the best known threshold of any qutrit magic state distillation routine. Our calculations rely on the geometric reformulation of magic state distillation in the language of discrete phase space, given in \cite{qudit-bound-states}. 

\section{Background}
\label{background}
While fault-tolerant quantum computing with qubits is now a widely-known subject, many aspects of fault-tolerant quantum computing with qudits of dimension other than two may be obscure to some readers. Therefore, in this section, we briefly summarize some necessary background, following \cite{nature}. This is a rich subject, and readers are encouraged to refer to some of the references cited below for a more thorough discussion.

\subsection{Qudit Pauli and Clifford Operators}

In this paper, we are interested in qutrits, which are quantum systems of dimension $d=3$. In this section, however, we present definitions which apply more generally to qudits of arbitrary odd prime dimension $d$. Let $\mathcal H_d$ be the Hilbert space for a single qudit. The computational basis for $\mathcal H_d$ consists of states $\ket{k}$, where $k$ is an element of the finite field $\mathbb Z_d$.

Generalized Pauli $X$ and $Z$ operators for qudits are defined as \cite{Gottesman1999}
\begin{equation}
    X\ket{k}=\ket{k+1}, ~ Z\ket{k}=\omega^k \ket{k},
\end{equation}
where $\omega=e^{2\pi i/d}$ is a $d$th root of unity. These operators satisfy $Z^d=X^d=1$ and $Z X=\omega X Z$. There are $d^2$ linearly independent Pauli operators, including the identity, which are also known as \textit{Heisenberg-Weyl displacement operators}, and are conventionally defined as\footnote{Here $2^{-1}$ is an element of the finite field $\mathbb Z_d$.} \cite{Gross},
\begin{equation}
 D_{(u|v)} = \omega^{2^{-1} uv}X^uZ^v.
\end{equation}
With this choice of overall phase, $D_{(u_1|v_1)}D_{(u_2|v_2)}=D_{(u_1+u_2|v_1+v_2)}$.  Multi-qudit Heisenberg-Weyl operators are defined as tensor products of single-qudit Heisenberg-Weyl operators,
 \begin{equation}
     D_{(\vec{u}|\vec{v})}=D_{(u_1,v_1)} \otimes \ldots D_{(u_n,v_n)}  
 \end{equation}
 and can be specified via a \textit{symplectic vector} $(\vec{u}|\vec{v})=(u_1, \ldots, u_n | v_1, \ldots, v_n)$.
 
 The Clifford group is defined as the set of unitaries that preserve Heisenberg-Weyl operators under conjugation.  Single-qudit Clifford unitaries act on Heisenberg-Weyl operators as  $SL(2,\mathbb Z_d)$ transformations. Explicitly, as shown in \cite{Appleby2}, any single-qudit Clifford unitary can be written in the form
 $D_{(u|v)}V_{ \hat{F} }$,
 where $\hat{F}$ is an element of $SL(2,\mathbb Z_d)$, i.e., a matrix $\begin{pmatrix} a & b \\ c & d \end{pmatrix}$, for some $a$, $b$, $c$ and $d \in \mathbb Z_d$ such that  $ad-bc = 1$. The operators $V_{\hat{F}}$ are known as symplectic rotations, and are given by the expression
\begin{equation}
V_{\hat{F}}=
\begin{cases}
\frac{1}{\sqrt{d}}  \sum_{j,k=0}^{d-1} \omega^{2^{-1} b^{-1}(ak^2-2jk+dj^2)}
\ket{j}\bra{k} & b\neq 0 \\
\sum_{k=0}^{d-1}\omega^{2^{-1} ack^2} \ket{ak}\bra{ k} & b= 0 
\end{cases}.
\end{equation}
Symplectic rotations act on Heisenberg-Weyl displacement operators as:
 \begin{equation}
     V_{\hat{F}} D_{(u|v)} V_{
     \hat{F}}^\dagger= D_{(u'|v')},
 \end{equation}
 where 
 \begin{equation}
     \begin{pmatrix} u' \\ v' \end{pmatrix} = \hat{F}
       \begin{pmatrix} u \\ v \end{pmatrix}.
 \end{equation}
 Upto an overall phase, they also satisfy  $V_{\hat{F}_1}V_{\hat{F}_2}=V_{\hat{F}_1 \hat{F}_2}$.
 
 In particular, it can be shown \cite{Appleby1, jain} that the Hadamard gate $H=V_{\hat{H}}$, acts on Pauli operators via $HXH^\dagger=Z$ and $HZH^\dagger=X^{-1}$, and therefore corresponds to the 
$SL(2,\mathbb Z_d)$ transformation,
\begin{equation}
    \hat{H}= \begin{pmatrix} 0 & -1 \\ 1 & 0 \end{pmatrix}. \label{HSL}
\end{equation}
Let us define another Clifford operator $N=V_{\hat{N}}$ to correspond to the $SL(2,\mathbb Z_d)$ transformation,
\begin{equation}
   \hat{N} = \begin{pmatrix} -1 & 0 \\ -1 & -1 \end{pmatrix}. \label{NSL}
\end{equation}
Explicitly, $N$ acts on qutrit Pauli operators via $NXN^\dagger =\omega^2 X^2Z^2$ and $NZN^\dagger=Z^2$. Together $\hat{N}$ and $\hat{H}$ generate all $SL(2,\mathbb Z_d)$ transformations, and the operators $ZN$ and $H$ can be shown to generate the entire single-qudit Clifford group. 
\subsection{Discrete Wigner Functions}
The Heisenberg-Weyl displacement operators are unitary but not Hermitian. A manifestly Hermitian basis for single-qudit density matrices is formed by the \textit{phase point operators} $A_{(u,v)}$, which are defined in terms of the Heisenberg-Weyl displacement operators as follows: 
\begin{eqnarray}
A_{(0,0)}&=&\frac{1}{d}\sum_{u=0}^{d-1}\sum_{v=0}^{d-1}  D_{(u,v)} \\ A_{(u,v)}&=&  D_{(u,v)}A_{(0,0)} D_{(u,v)}^\dagger. \label{phase-point-def}
\end{eqnarray}
Any qudit density matrix $\rho$ can be expressed as a linear combination of phase-point operators with real, but possibly negative, coefficients, 
\begin{equation}
W_\rho(u,v)=\frac{1}{d} \text{Tr}\left(\rho A_{(u,v)}\right).
\end{equation}
This representation, which completely characterizes the quantum state $\rho$, is known as its \textit{discrete Wigner function}, and is the natural generalization of the well-known continuous quasi-probability distribution introduced by Wigner \cite{wigner1932} for finite dimensional systems. It was first introduced in \cite{Wootters1987} and further developed in \cite{PhysRevA.70.062101, cormick2006classicality, Gross,  Appleby2}. The Wigner function for an  $n$-qudit state can be defined analogously using tensor products of the phase-point operators \cite{Veitch_2012}. If the $n$-qudit state is separable, its Wigner function can be written as a product of single-qudit Wigner functions.

The convex subset of state space with non-negative discrete Wigner functions is known as the \textit{Wigner polytope}. An $n$-qudit quantum state within the Wigner polytope can be thought of as a probability distribution over ontological states $(\vec{u},\vec{v}) \in (\mathbb Z_d \otimes \mathbb Z_d)^n$, defined by the phase point operators, known as \textit{discrete phase space}. From the definition \eqref{phase-point-def}, it is clear that qudit Pauli operators and Clifford unitaries act as discrete translations and symplectic rotations on phase space. The action of these operators on multi-qudit states with (efficiently sampleable) non-negative Wigner functions can therefore be efficiently simulated, via a Monte-Carlo type approach, as explained in more detail in \cite{Veitch_2012}.

Because Clifford operations and stabilizer measurements on states within the Wigner polytope can be efficiently simulated, these states are not useful for achieving universal quantum computation via state-injection. The Wigner polytope therefore provides a bound for the threshold of any magic state distillation routine, much like the stabilizer polytope for qubits.\footnote{We should point out that, as observed in \cite{Veitch_2012}, for qudits of odd-prime dimension, the stabilizer polytope is a proper subset of the Wigner polytope.} 
Negativity of the Wigner function can therefore be thought of as a resource for quantum computation, in a sense that is made precise in \cite{Veitch_2012}. Wigner negativity also turns out to be equivalent to contextuality, as shown in \cite{nature}.

\subsection{Discrete Phase Space Formulation of Qudit Magic State Distillation} 
Qudit magic state distillation was recast in the language of discrete phase space in \cite{qudit-bound-states}. This formulation of magic state distillation is particularly convenient to implement computationally, and we will use it to determine the performance of the 11-qutrit ternary Golay code for magic state distillation below. Let us briefly review it here.

A magic state distillation routine takes as input $n$ noisy qudits, which are in the state $\rho_{\text{in}}\otimes \rho_{\text{in}}\otimes \ldots \otimes \rho_{\text{in}}$, and produces a single higher-fidelity qudit in the state $\rho_{\text{out}}$. The routine consists of first projecting the input qudits onto the codespace of an $n$-qudit stabilizer code, which is described by an $(n-1)\times 2n$ symplectic matrix $\mathbf{M}$; and then decoding the resulting state to obtain a single output qudit. The decoding step depends on the choice of logical operators for the code, $\bar{X}$ and $\bar{Z}$, which can be specified by the symplectic vectors $(\vec{a}_x |~\vec{b}_x)$ and $(\vec{a}_z |~\vec{b}_z)$. 

The general idea behind \cite{qudit-bound-states} is as follows. The inverse of a magic state distillation routine is an encoding circuit for the stabilizer code, that can be thought of as a linear map (an isometry) from the logical Hilbert space $\mathcal H_d$ to the physical Hilbert space $\mathcal H_d^n$. In the language of discrete phase space, this translates to a multi-valued function $\mathcal E: \mathbb Z_d\otimes \mathbb Z_d \rightarrow (\mathbb Z_d\otimes \mathbb Z_d)^n$, from logical phase space to physical phase space. The image of a point $(z_L, x_L) \in \mathbb Z_d\otimes \mathbb Z_d$ under $\mathcal E$ consists of $d^{n-1}$ points in $(\mathbb Z_d\otimes \mathbb Z_d)^n$, determined explicitly in \cite{qudit-bound-states}. The Wigner function of the decoded logical qudit at $(z_L, x_L)$ is simply the sum of the Wigner function of the physical qudits, evaluated at each of these $d^{n-1}$ points (up to an overall normalization constant.)  

The explicit expression for the Wigner function, $W_{\text{out}}(z,x)$, corresponding to the single-qudit density matrix $\rho_{\text{out}}$, in terms of the Wigner function, $W_{\text{in}}(z,x)$, corresponding to single-qudit density matrix $\rho_{\text{in}}$ is
\begin{equation}
W_{\text{out}}(z_L,x_L) = \frac{1}{\mathcal P}\sum_{\vec{u} \in \mathbb{Z}_d^{n-1}} \prod_{i=1}^{n} W_{\text{in}}(z_i(\vec{u},z_L,x_L),x_i(\vec{u},z_L,x_L) ). \label{algorithm}
\end{equation}
Here, $\mathcal P$ is the probability for successful projection onto the stabilizer code, and is determined by the condition that $$\sum_{z_L,x_L} W_{\text{out}}(z_L,x_L)=1.$$
The quantities $z_i$ and $x_i$ are the $i$th components of the vectors $\vec{x}$ and $\vec{z}$ given by,
\begin{equation}
\begin{pmatrix} \vec{z}(\vec{u},z_L,x_L) \\ \vec{x}(\vec{u},z_L,x_L) \end{pmatrix} = \begin{pmatrix}\mathbf{M}^T & \begin{matrix}\vec{a}_z  \\ \vec{b}_z \end{matrix} &  \begin{matrix}\vec{a}_x  \\ \vec{b}_x \end{matrix} 
\end{pmatrix} \begin{pmatrix}\vec{u} \\ -z_L \\ -x_L \end{pmatrix}. 
\end{equation}

We checked that this algorithm is able to reproduce the results of previous studies of qutrit magic state distillation, such as \cite{ACB}.

\section{The strange state and the Norell state}

The strange state, 
\begin{equation}
    \ket{S} = \frac{1}{\sqrt{2}} \left(\ket{1} - \ket{2}\right),
\end{equation}
and the Norell state,
\begin{equation}
    \ket{N} = \frac{1}{\sqrt{2}} \left(\ket{1} + \ket{2}\right),
\end{equation}
are both eigenstates of the single-qutrit Clifford operator $N$, defined as,
\begin{equation}
   N = \begin{pmatrix} 1 & 0 & 0 \\ 0 & 0 & \omega^2 \\ 
    0 & \omega^2 & 0 \end{pmatrix}.
\end{equation}
The third eigenvector of $N$ is $\ket{0}$.

The strange state is also an eigenvector of the qutrit Hadamard gate,
\begin{equation}
    H =\frac{1}{\sqrt{3}} \begin{pmatrix} 
    1 & 1 & 1 \\
    1 & \omega & \omega^2 \\
    1 & \omega^2 & \omega 
    \end{pmatrix}
\end{equation}
with eigenvalue $i$. The other two eigenstates of $H$ are $\ket{H_1}$ and $\ket{H_{-1}}$, with eigenvalues $+1$ and $-1$. These are given by
\begin{equation}
    \ket{H_{\pm 1}} = \cos \phi \ket{0} \pm \frac{1}{\sqrt{2}}\sin \phi(\ket{1}+\ket{2}),
\end{equation}
where $\phi=\frac{1}{2} \arctan \sqrt{2}$. 

\begin{figure}
    \centering
    \includegraphics[scale=.2]{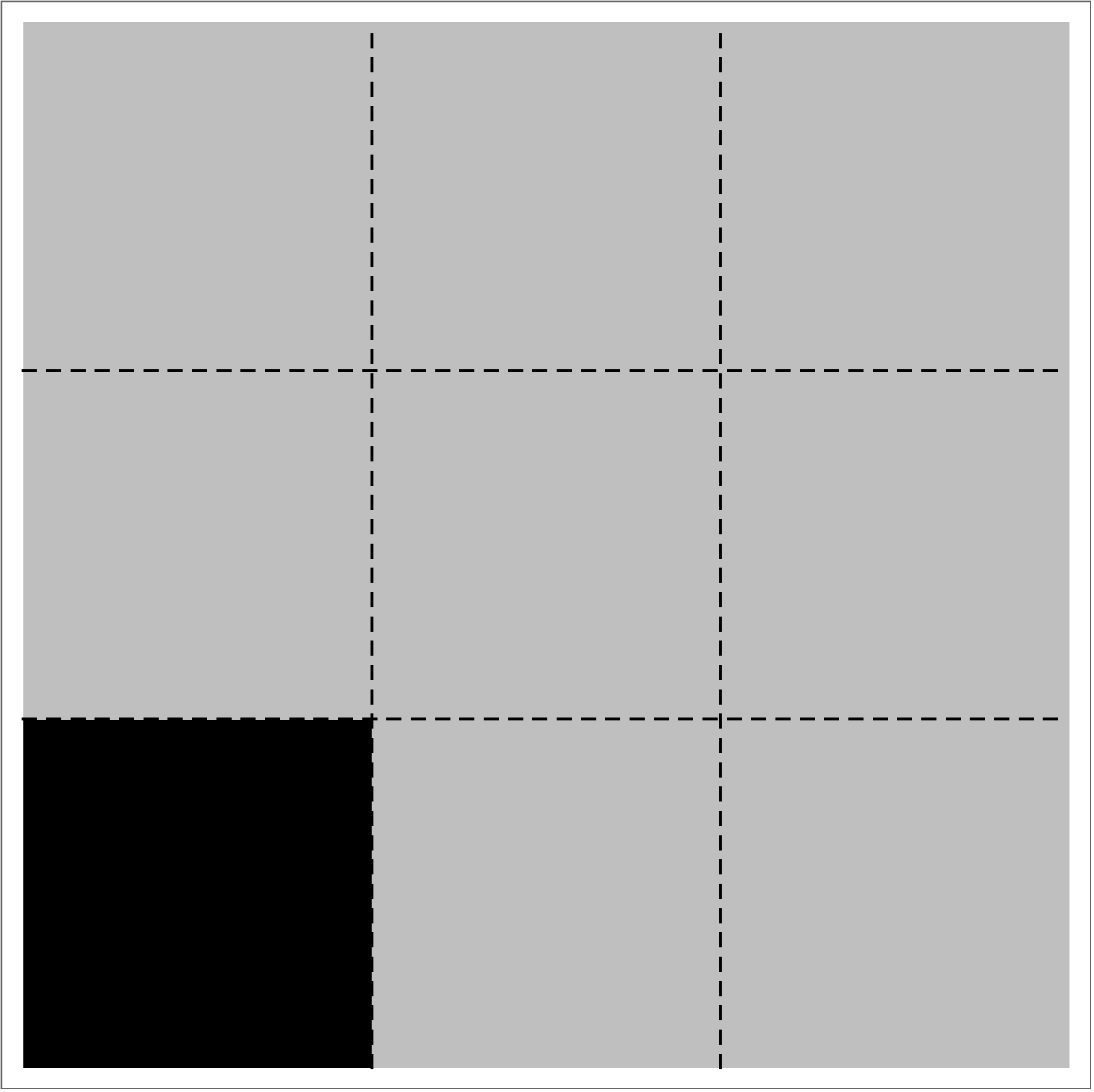} $~$
    \includegraphics[scale=.2]{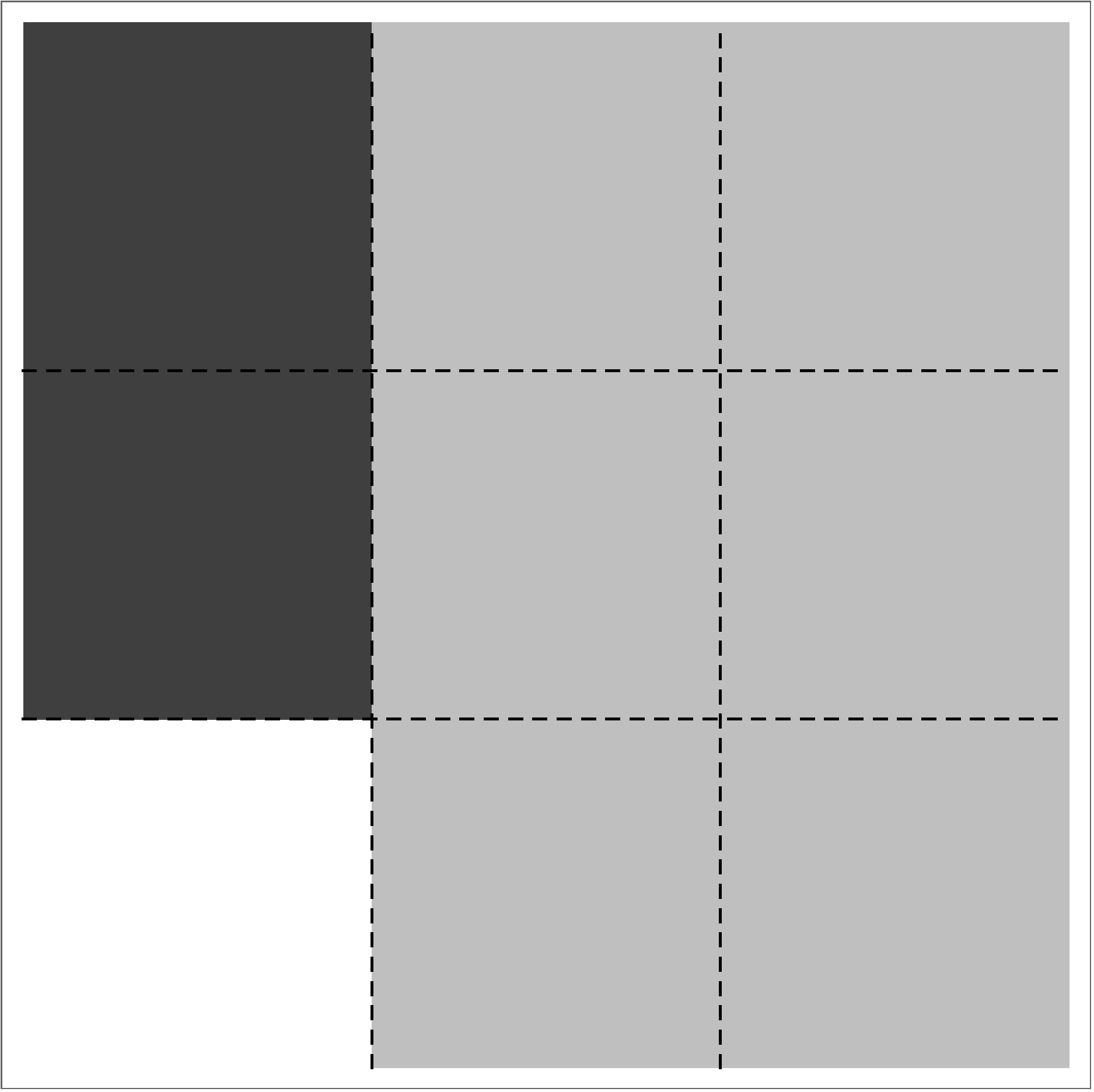}
    \includegraphics[scale=.32]{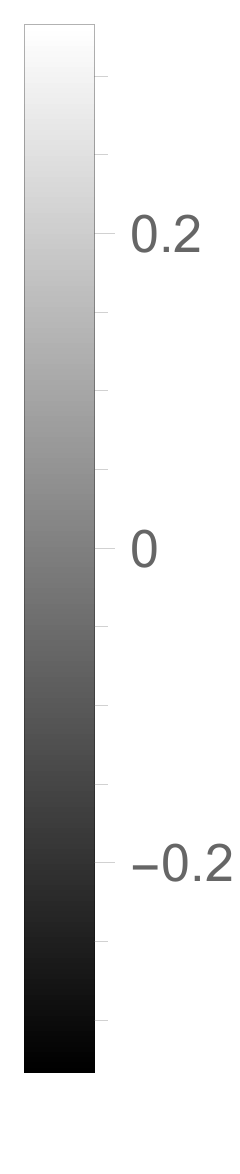}
    \caption{The discrete Wigner function for the strange state (left) and the Norell state (right), obtained from eqs. \eqref{discrete-wigner-noisy} and  \eqref{discrete-wigner-norell}. In this figure, $W_\rho(u,v)$ is plotted on a $3\times3$ grid, with $u$ on the horizontal axis and $v$ on the vertical axis. The point $(u,v)=(0,0)$ corresponds to the square in the bottom left corner, and $(u,v)=(2,2)$ corresponds to the square in the top right corner.}
    \label{fig:wigner}
\end{figure}

The discrete Wigner functions for $\ket{S}$ and $\ket{N}$ are plotted in Figure \ref{fig:wigner}.  The symmetries of these and other qutrit Clifford eigenstates are discussed in detail in \cite{jain}. Any of these states can be used to implement a non-Clifford gate via state injection, as we review in Appendix A, which closely follows \cite{ACB}.

In the magic state model, we will begin with a supply of noisy $\ket{S}$ and $\ket{N}$ states, that lie somewhere near $\ket{S}$ or $\ket{N}$ in the 8-dimensional space of single-qutrit density matrices. Via random application of Clifford unitaries, a process known as \textit{twirling}, one can restrict the density matrices of noisy input qutrits to a more manageable form. While the twirling schemes for qubits in \cite{MSD} result in a one-parameter family of density matrices; for qutrits, we generically expect a twirling scheme to result in a two-parameter family of density matrices.

For noisy Norell states,   apply the unitary $N^n$, where $n$ is a random integer between $0$ and $5$, to define the following map:
\begin{equation}
    \rho \rightarrow \frac{1}{6} \sum_{n=0}^5 N^{n}\rho N^{-n},
\end{equation}
This restricts our noisy states to lie in the two-dimensional plane spanned by convex combinations of the three eigenvectors of $N$: $\ket{0}$, $\ket{N}$ and $\ket{S}$, \begin{equation}
    \rho_N(\epsilon_0,\epsilon_S) = (1-\epsilon_0-\epsilon_S)\ket{N}\bra{N}+\epsilon_0 \ket{0}\bra{0} + \epsilon_S\ket{S}\bra{S}, \label{norell-states}
\end{equation}
where $\epsilon_0=\bra{0}\rho\ket{0}$ and $\epsilon_S=\bra{S}\rho \ket{S}$.
The space of density matrices parameterized by equation \eqref{norell-states} forms an equilateral triangle, and is pictured in Figure \ref{triangle} below.

The strange state $\ket{S}$ is the unique simultaneous eigenstate of two Clifford unitaries $H$ and $N$. These two elements generate a subgroup of the Clifford group isomorphic to $SL(2, \mathbb Z_3)$, described in Section \ref{background}. As observed in \cite{jain}, by randomly choosing to apply any element of this finite group, any noisy input state can be brought into the form,
\begin{equation}
    \rho_S(\delta)=(1-\delta)\ket{S}\bra{S} + \delta \frac{ 1}{3}. \label{noisy-S-states}
\end{equation}

To see this explicitly, first apply the operator $H$ to the noisy input state $n$ times, where $n$ is a random integer between $0$ and $3$, to define the map:
\begin{equation}
    \rho \rightarrow \frac{1}{4} \sum_{n=0}^3 H^n \rho H^{-n}.
\end{equation}
The resulting density matrix must then be expressible as a mixture of the eigenstates of $H$, 
\begin{equation}
    \rho(\epsilon_1, \epsilon_2) = (1-\epsilon_1-\epsilon_2)\ket{S}\bra{S} + \epsilon_1 \ket{H_{1}}\bra{H_{1}} + \epsilon_2 \ket{H_{-1}}\bra{H_{-1}}, \label{noisy-S-states-2}
\end{equation} 
where $\epsilon_1=\bra{H_{1}}\rho \ket{H_1}$ and $\epsilon_2=\bra{H_{-1}}\rho \ket{H_{-1}}$.
It can be shown \cite{jain} that the unitary operator $$H'=-i\ket{S}\bra{S} - e^{i\pi/4}\ket{H_1}\bra{H_{-1}} + e^{3i\pi/4}\ket{H_{-1}}\bra{H_1}$$ is an element of the Clifford group.\footnote{One can check that $H'$ can be written as $V_{\hat{H}'}$, corresponding to the $SL(2,\mathbb Z_3)$ transformation $\hat{H}'=\begin{pmatrix} 1 & 1 \\ 1 & 2 \end{pmatrix}$.} $H'$ interchanges $\ket{H_+}$ and $\ket{H_-}$, but preserves $\ket{S}$. By randomly choosing whether or not to apply $H'$ to noisy input qutrits in the state $\rho(\epsilon_1,\epsilon_2)$,
\begin{equation}
    \rho(\epsilon_1,\epsilon_2) \rightarrow \frac{1}{2}\rho(\epsilon_1,\epsilon_2) + \frac{1}{2} H' \rho(\epsilon_1, \epsilon_2) (H')^{\dagger}=\rho(\epsilon/2,\epsilon/2),
\end{equation}
we obtain a state with $\epsilon_1 = \epsilon_2 \equiv \epsilon/2$.

It is easy to see that this density matrix is equivalent to \eqref{noisy-S-states}. We choose to express it in terms of the parameter $\delta=3\epsilon/2$, which can be interpretted as the depolarizing noise rate. After twirling, our $n$ noisy input qutrits are in the state $\rho_{\text{in}}^{\otimes n}=\rho_S(\delta_{\text{in}})^{\otimes n}$. Assuming the stabilizer code employed for distillation has suitable symmetries, the distilled output qutrit will be in a state of the same form $\rho_{\text{out}}=\rho_S(\delta_{\text{out}})$, thus giving rise to single a function of one variable $\delta_{\text{out}}(\delta_{\text{in}})$ that characterizes its performance, much like the qubit case.

The existence of a twirling protocol that converts all noise to depolarizing noise is a unique feature of the $\ket{S}$ state, that arises because of its exceptional symmetry properties under Clifford transformations. \cite{jain} This property is neither shared by any other qutrit magic state nor is it expected to hold for any other qudit magic state, for any odd prime $d>3$. 

\section{The 11-qutrit Golay code}

Consider any maximal self-orthogonal\footnote{A code $C$ is said to be self-orthogonal if $C \subseteq C^\perp$. A self-orthogonal  code $C$ is said to be \textit{maximal} if $C$ is not contained in any other self-orthogonal code. Ternary maximal self-orthogonal codes have been studied extensively in the literature, see, e.g., \cite{golayUnique, selfdual1, selfdual2}. The ternary Golay code itself is not self-orthogonal, but its dual  is.} classical ternary code \cite{selfdual1} of odd length, with generator matrix ${\mathbf M}_c$. We construct a quantum error correcting code from two copies of $\mathbf M_c$, following the CSS construction \cite{CSS1,CSS2}, with the following symplectic matrix:
\begin{equation}
    \mathbf{M}_{q}=\begin{pmatrix}
     \mathbf{M}_c &|& 0 \\ 0 &|& \mathbf{M}_c
    \end{pmatrix}. \label{CSS-code}
\end{equation}
It can be shown (e.g., \cite{golayUnique, selfdual1}), that any maximal self-orthogonal ternary code of odd length $n$,  has dimension $k=(n-1)/2$. The quantum code generated by this construction therefore encodes one qutrit. 

From equation \eqref{HSL}, we see that acting on such a code with $H^{\otimes n}$, results in a stabilizer code described by the symplectic matrix,
\begin{equation}
    \mathbf{M}_{q}'=\begin{pmatrix}
     0 &|& \mathbf{M}_c  \\  -\mathbf{M}_c &|& 0
    \end{pmatrix},
\end{equation}
which is clearly equivalent to $\mathbf{M}_{q}$. Similarly, from equation \eqref{NSL}, we see that acting on the code with $N^{\otimes n}$, results in a stabilizer code described by the symplectic matrix,
\begin{equation}
    \mathbf{M}_{q}''=\begin{pmatrix}
     -\mathbf{M}_c &|& -\mathbf{M}_c  \\
     0 &|& -\mathbf{M}_c 
    \end{pmatrix},
\end{equation}
which is also equivalent to $\mathbf{M}_{q}$. Therefore, the projector onto the stabilizer code described by $\mathbf{M}_{q}$ commutes both $H^{\otimes n}$ and $N^{\otimes n}$.

Let us choose for $\mathbf{M}_c$ the generator matrix for the dual of the ternary Golay code, which is self-orthogonal,
\begin{equation}
    \mathbf{M}_{cG} = 
     \left(
\begin{array}{rrrrrrrrrrr}
 -1 & 1 & 1 & -1 & -1 & 0 & 1 & 0 &
   0 & 0 & 0 \\
 -1 & 1 & -1 & 1 & 0 & -1 & 0 & 1 &
   0 & 0 & 0 \\
 -1 & -1 & 1 & 0 & 1 & -1 & 0 & 0 &
   1 & 0 & 0 \\
 -1 & -1 & 0 & 1 & -1 & 1 & 0 & 0 &
   0 & 1 & 0 \\
 -1 & 0 & -1 & -1 & 1 & 1 & 0 & 0 &
   0 & 0 & 1 \\
\end{array}
\right). \label{golay}
\end{equation}
The Golay code is of length $11 \equiv 2 \mod 3$.  $X^{\otimes n}$ and $Z^{\otimes n}$ cannot be stabilizers of this code since the classical ternary vector $(1, \ldots, 1)$ is not self-orthogonal. However, both $X^{\otimes n}$ and $Z^{\otimes n}$ can serve as logical Pauli operators for the code. Let us denote logical operators and states with an overbar. We make the choice
\begin{equation}
    \bar{X} = (X)^{\otimes n},~\bar{Z}^\dagger=Z^{\otimes n}.
\end{equation}
With the above choice, the logical $H$ and $N$ operators are given by,
\begin{equation}
  \bar{H} = (H^\dagger)^{\otimes n},~\bar{N}=(N^\dagger)^{\otimes n}.
\end{equation} Together with $\bar{X}$ and $\bar{Z}$, these form a complete set of transversal Clifford gates. 

Recall that Bravyi and Kitaev defined two qubit magic states in \cite{MSD}: $\ket{H}$ and $\ket{T}$. Stabilizer codes used to distill $\ket{H}$ states, such as the 15-qubit code of \cite{MSD} and the codes in \cite{Reichardt2005, bravyi2012magic}, crucially support a non-Clifford transversal gate. The successful distillation of $\ket{H}$ states by these codes can be understood as a direct consequence of the existence of this transversal gate. No such understanding is available for distillation of $\ket{T}$ states via the 5-qubit code \cite{MSD}, which does not support any transversal gate outside the Clifford group. For the 5-qubit code, distillation must be demonstrated by what is essentially a brute-force calculation of projection of noisy input states onto the stabilizer code. 

The 11-qutrit Golay code supports a complete set of transversal Clifford gates  -- therefore, by the Eastin-Knill theorem \cite{EastinKnill}, it cannot support a non-Clifford transversal gate. For this reason, distillation via the 11-qutrit Golay code is analogous to distillation via the 5-qubit code, where the mechanism for distillation is somewhat opaque.

\begin{figure}
    \centering
    \includegraphics[scale=.3]{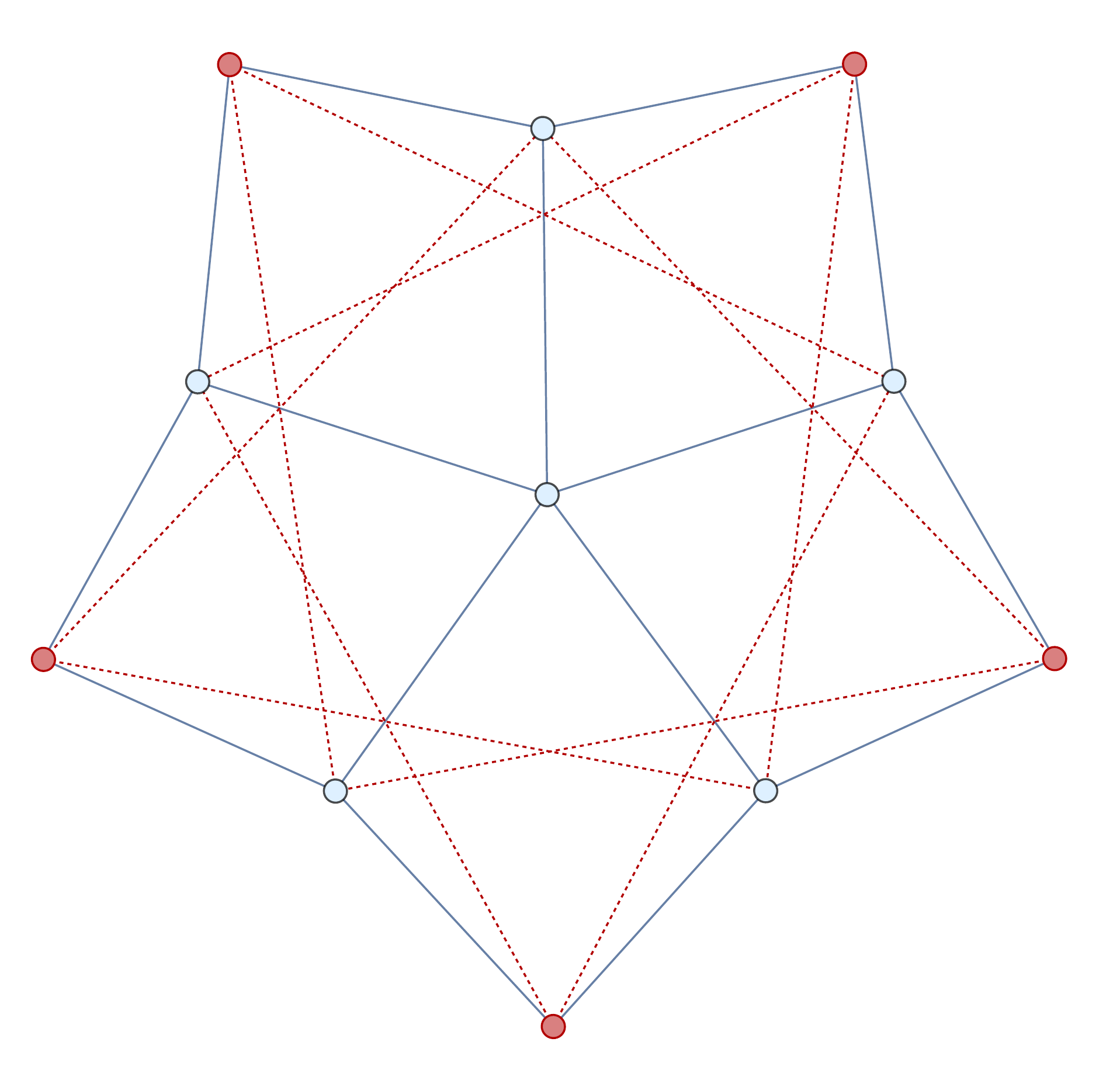}
    \caption{A graph state representation of the 11-qutrit Golay code. By applying a sequence of local Clifford unitaries and elementary row operations described in \cite{GF9}, the symplectic matrix for the stabilizers of the logical $\ket{\bar{0}}$ state can be brought into the form $(\mathbf 1 | \Gamma)$ where $\Gamma$ is the adjacency matrix of an undirected graph whose edges have weights in $\mathbb Z_3$, pictured above. Dotted red edges have weight $1$ and blue edges have weight $-1$. Vertices correspond to qutrits, and a subset of the vertices are highlighted in red. The tensor product of Pauli-$Z$ operators acting on each of the highlighted vertices defines the logical $\bar{X}$ operator, as in \cite{CWS}.}
    \label{graph-state}
\end{figure}

Let us conclude this section by pointing out that the 11-qutrit Golay code can be represented as a graph state \cite{Schlingemann2002StabilizerCC}, in the spirit of \cite{CWS}, following the procedure given in \cite{GF9}. This is shown in Figure \ref{graph-state}.

\section{Distilling the strange state}

Following \cite{MSD}, a natural requirement for a code to be suitable for distillation of $\ket{S}$ states is that $\ket{S}^{\otimes n}$ decode to $\ket{\bar{S}}$ after projection onto the codespace. Let us see that this is indeed the case for the ternary Golay code. Let $\Pi$ be the projector onto the codespace. We require $\Pi \ket{S}^{\otimes n} \propto \ket{\bar{S}}$.
Since $\Pi$ commutes with $\bar{H}$,  we have
\begin{equation}
    \bar{H} \left(\Pi \ket{S}^{\otimes n}
    \right)= \Pi \bar{H}\ket{S}^{\otimes n} = i^{-n} \Pi \ket{S}^{\otimes n}. 
\end{equation}
We see that, for $n=11$, $\Pi \ket{S}^{\otimes n}$ is an eigenvector of $\bar{H}$ with eigenvalue $i$, and therefore must be proportional to $\ket{\bar{S}}$. The coefficient of proportionality determines the probability of successfully projecting onto the code, and it remains to demonstrate that this probability is nonzero -- i.e., that $\Pi \ket{S}^{\otimes n}\neq0$. This requires a more non-trivial computation, which we carry out next. 

Before proceeding, note that, our analysis so far has been very general, and applies to a CSS code constructed from two copies of any maximal self-orthogonal ternary code of length $n=12m-1$, with the property that the ternary vector $(1~1~\ldots~1)$ is orthogonal to all its generators. There are three such codes of length $11$, given in \cite{selfdual1}. Using such a code for distillation of $\ket{S}$ states, we generically expect the noise rate of the distilled qutrit to depend \textit{linearly} on the noise rate of the input qutrits, for the following reason. 

The eigenstates of the qutrit Hadamard operator are $\ket{S}$, $\ket{H_{1}}$ and $\ket{H_{-1}}$, with eigenvalues $i$, $1$ and $-1$, respectively. Let us denote $\ket{H_0}\equiv\ket{S}$, so that we can define
\begin{equation}
    \ket{H_{\vec{x}}} \equiv \ket{H_{x_1}}\otimes \ket{H_{x_2}}\otimes\ldots \otimes \ket{H_{x_n}}
\end{equation}
which depends on the ternary string $\vec{x}=(x_1, \ldots, x_n)$, each of whose entries are $0$, $1$ or $-1$. The density matrix for $n$ noisy strange states, each described by $\rho(\epsilon/2,\epsilon/2)$ given in equation \eqref{noisy-S-states-2}, can be written as,
\begin{equation}
    \rho_{S}^{\otimes n} = \sum_{\vec{x} \in \{0,1,-1\}^{n}} \left(\frac{\epsilon}{2}\right)^{|\vec{x}|}(1-\epsilon)^{n-|\vec{x}|} \ket{H_{\vec{x}}}\bra{H_{\vec{x}}}.
\end{equation}
Here, $|\vec{x}|$ is the weight (number of nonzero entries) of the ternary vector $\vec{x}$.
The (unnormalized) density matrix for the output qutrit is
\begin{equation}
    \tilde{\rho}_{\text{out}} =  \sum_{\vec{x} \in \{0,1,-1\}^{n}} \left(\frac{\epsilon}{2}\right)^{|\vec{x}|}(1-\epsilon)^{n-|\vec{x}|} \Pi \ket{H_{\vec{x}}}\bra{H_{\vec{x}}}\Pi^\dagger. \label{un-normalized}
\end{equation}
For any $\vec{x}$, $\Pi \ket{H_{\vec{x}}}$ is an eigenvector of $\bar{H}$, and is proportional to one of $\ket{\bar{S}}$, $\ket{\bar{H}_1}$ or $\ket{\bar{H}_{-1}}$ (unless it vanishes). 
After normalization, the output density matrix can therefore be written as\footnote{Here we are also using the fact that $(H'
)^{\otimes n}$ defined earlier commutes with the codespace, maintaining the symmetry between $\ket{H_{1}}$ and $\ket{H_{-1}}$.}
\begin{equation}
    \rho_{\text{out}} =  (1-\epsilon_{\text{out}})\ket{\bar{S}}\bra{\bar{S}} + \frac{\epsilon_{\text{out}}}{2}\left(\ket{\bar{H}_1}\bra{\bar{H}_1} + \ket{\bar{H}_{-1}}\bra{\bar{H}_{-1}}\right)
\end{equation}
where $\epsilon_{\text{out}}$ is some function of $\epsilon$. 

We saw above that, when $|\vec{x}|=0$, $\Pi \ket{H_{\vec{x}}}$ is proportional to $\ket{\bar{S}}$. Let us look next at the term proportional to $\frac{\epsilon}{2}(1-\epsilon)^{n-1}$, i.e., states $\Pi \ket{H_{\vec{x}}}$ for which $|\vec{x}|=1$. Any such state is an eigenvector of $\bar{H}$ with eigenvalue $\pm 1$. Unless each such state happens to be perfectly orthogonal to the codespace, it will decode to one of the logical states $\ket{\bar{H}_{\pm 1}}$, resulting in a contribution to $\epsilon_{\text{out}}$ linear in $\epsilon$.

Remarkably, it turns out that the $|\vec{x}|=1$ term is indeed perfectly orthogonal to the codespace for the 11-qutrit Golay code and therefore the linear contribution to $\epsilon_{\text{out}}$ vanishes. Note that, if the linear contribution to $\epsilon_{\text{out}}$ vanishes, the next possible contribution is cubic in $\epsilon$. The reason for this is that, when $|\vec{x}|=2$, any term of the form $\Pi \ket{H_{\vec{x}}}$ must be an eigenvector of $\bar{H}$ with eigenvalue $\pm i$. Since $\bar{H}$ has no eigenvector with eigenvalue $-i$, terms of this form must either vanish, or be proportional to $\ket{\bar{S}}$. Terms with $|\vec{x}|=2$, therefore, do not contribute to $\epsilon_{\text{out}}.$

We use the algorithm of \cite{qudit-bound-states}, reviewed in section \ref{background}, to simulate both projection onto the stabilizer code and subsequent decoding. The discrete Wigner function corresponding to $\rho(\delta)$ in equation \eqref{noisy-S-states} is:
\begin{equation}
W(u,v;\delta) = \begin{cases} -\frac{1}{3} + \frac{4}{9}\delta & (u,v)=(0,0) \\ \frac{1}{6} - \frac{1}{18}\delta & (u,v) \neq (0,0)\end{cases}.
\label{discrete-wigner-noisy}
\end{equation}
This is shown in Figure \ref{fig:wigner}, for $\delta=0$. We used a computer algebra system (Mathematica 12) to  evaluate $W_{\text{out}}(u,v)$ using equation \eqref{algorithm}, with  $W_{\text{in}}=W(u,v;\delta)$. (The Mathematica notebook is included as electronic supplementary information.) 

\begin{figure}
    \centering
    \includegraphics[width=.6\textwidth]{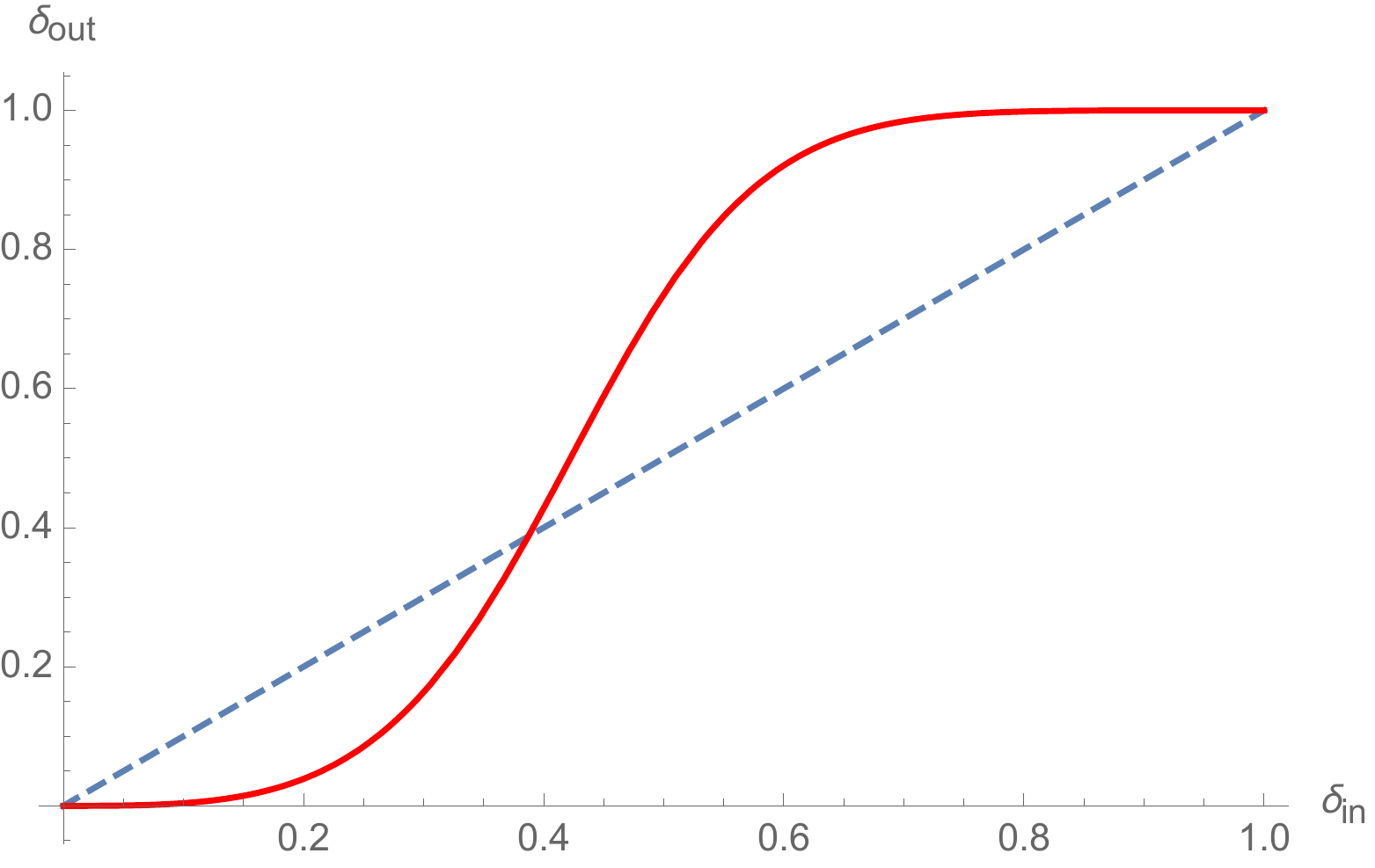}
    \caption{The relation $\delta_{\text{out}}(\delta_{\text{in}})$ induced by distillation with the 11-qutrit Golay code is shown by the solid red line. The dashed line is the line $\delta_{\text{out}}=\delta_{\text{in}}$, which is shown for convenience. If $\delta$ is below the threshold value of $0.387$, where both lines intersect, the noise rate of the output qutrit is less than that of the input qutrits, i.e., $\delta_{\text{out}}<\delta_{\text{in}}$. }
    \label{fig:golay}
\end{figure}

As expected, $W_{\text{out}}(u,v)$ is of the form $W(u,v;\delta_{\text{out}})$, with $\delta_{\text{out}}(\delta)$ given by:
\begin{equation}
    \delta_{\text{out}} = \delta^3
    \frac{P(\delta)}{2Q(\delta)},
\end{equation}
where
\begin{eqnarray}
P(\delta) & = & 3021 \delta ^8-24816 \delta ^7+92180 \delta ^6-203280 \delta ^5+292710 \delta ^4-283536 \delta ^3+181764 \delta ^2 \nonumber \\ && -71280 \delta +13365 \\
Q(\delta) & = & 495 \delta ^{11}-3960 \delta ^{10}+13750 \delta ^9-25245 \delta ^8+18810 \delta ^7+23628 \delta ^6-86328 \delta ^5 \nonumber\\
&& +121770 \delta ^4-102465 \delta ^3+53460 \delta ^2-16038 \delta +2187
\end{eqnarray}
This is plotted in Figure \ref{fig:golay}.

For small $\delta$, 
\begin{equation}
    \delta_{\text{out}} \approx \frac{55}{18} \delta^3.
\end{equation}
As mentioned above, for qutrit magic state distillation routines, we generically expect a linear relation between $\delta_{\text{out}}$ and $\delta$; so this cubic noise suppression is fairly surprising. 
Our derivation of this result is, essentially, computational. We hope  to better understand the origin of this cubic rate of error-suppression theoretically, perhaps as a consequence of the symmetries of the ternary Golay code; but this is beyond the scope of the present work.

The threshold for distillation is at \begin{equation}
    \delta_*=\frac{3}{135} \left(31-262
   \sqrt[3]{\frac{2}{405
   \sqrt{109}-2981}}+2^{2/3}
   \sqrt[3]{405
   \sqrt{109}-2981}\right) \approx 0.38715
\end{equation}
This is slightly more than half of the theoretical upper bound for the threshold determined by the Wigner polytope \cite{qudit-bound-states, nature, Veitch_2012}, which is at $\delta_* = \frac{3}{4}$.
This threshold is better than the best previously known threshold for any qutrit magic state distillation protocol. (The best previously known threshold to depolarizing noise was achieved by a distillation routine in \cite{Howard} that had only linear error-suppression.)

The probability $\mathcal P$ for successful distillation is the trace of the unnormalized density matrix $\tilde{\rho}_{\text{out}}$ in equation \eqref{un-normalized}, and is given by,
\begin{equation}
    \mathcal P=\frac{1}{1728}-\frac{11  }{2592} \delta +\frac{55  ^2}{3888}\delta+O\left(\delta ^3\right).
\end{equation}
 The low success rate means that, in practice, approximately $19008$ qutrits would be needed for a single successful round of distillation. This is offset slightly by the cubic error suppression, which implies that, starting with $n$ noisy copies of the strange state with depolarizing noise rate $\delta$, the noise rate of the distilled strange state scales with $n$ as
\begin{equation}
\delta_{\text{out}}(n,\delta) \approx \frac{1}{1.75}\left( 1.75 \delta \right)^{n^{0.112}}
\end{equation}
where $\xi=\frac{1}{\log_3 19008} \approx .112$ is the \textit{yield parameter}.  For comparison, with the 5-qubit code \cite{MSD}, we obtain a similar relation,
\begin{equation}
\delta^{5-\text{qubit}}_{\text{out}}(n,\delta) \approx \frac{1}{2.5}\left( 2.5 \delta \right)^{n^{0.204}}
\end{equation}
with yield parameter $\xi_{5-\text{qubit}}= \frac{1}{\log_2 30} \approx .204$.

\section{Distilling  the Norell state}
The ternary Golay code can also be used to distill  Norell states. While the Norell state is less magic than the strange state, if we restrict our operations to two-qutrit stabilizer measurements and Clifford unitaries, the Norell state is slightly more useful for state injection, as discussed in Appendix A.

\begin{figure}
    \centering
    \includegraphics[scale =.8]{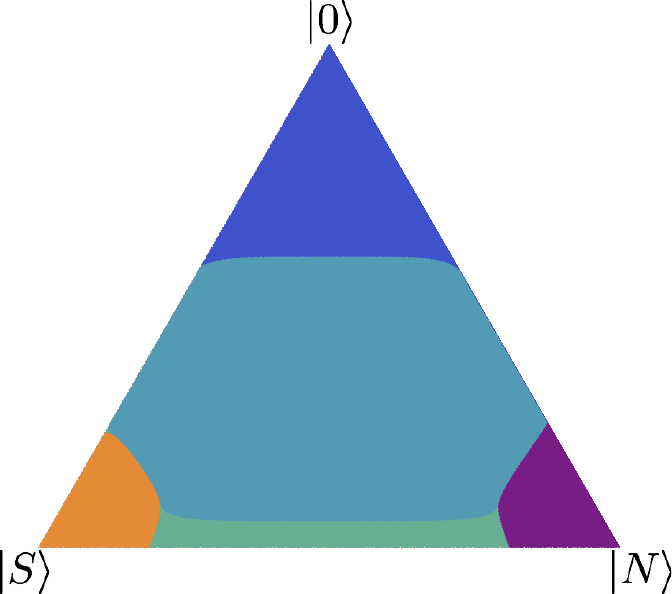}
    \caption{By randomly applying the Clifford operator $N$, any state can be made to lie in the triangle spanned by convex combinations of $\ket{S}$, $\ket{N}$ and $\ket{0}$. The purple region distills to the Norell state. The orange region distills to $\ket{S}$ and the dark blue region distills to $\ket{0}$. The teal and green regions distill to mixed states.}
    \label{triangle}
\end{figure}

After twirling, noisy Norell states are described by the density matrix $\rho_N(\epsilon_0, \epsilon_S)$, given in \eqref{norell-states}. This corresponds to the discrete Wigner function:
\begin{equation}
W(u,v;\delta) = \begin{cases} \frac{1}{3} (1-2 \epsilon_S) & u=0,~v=0 \\ \frac{1}{6} (3 \epsilon_0+2 \epsilon_S -1) & u=0,~v \neq 0 \\
\frac{1-\epsilon_0}{6} & u \neq 0\end{cases}.
\label{discrete-wigner-norell}
\end{equation}
Our distillation routine takes 11 qutrits in the state $\rho(\epsilon_0,\epsilon_S)^{\otimes 11}$ and outputs a single qutrit in the state $\rho(\epsilon_0',\epsilon_S')$, and is thus characterized by the two functions $\epsilon_0'(\epsilon_0, \epsilon_S)$, and  $\epsilon_S'(\epsilon_0,\epsilon_S))$. We obtained these expressions, which are presented in Appendix B, using the simulation algorithm of \cite{qudit-bound-states}, as in the previous section. For small $\epsilon_0$ and $\epsilon_S$, these come out to be:
\begin{eqnarray}
\epsilon_0' & = & \epsilon_0^2 \left(\frac{55}{18}+\frac{55 \epsilon_S}{9}+\frac{715 \epsilon_S^2}{6}+O\left(\epsilon_S^3\right)\right)+O\left(\epsilon_0^3\right) \\
\epsilon_S' & = & \left(\frac{55 \epsilon_S^3}{3}+O\left(\epsilon_S^4\right)\right)+\epsilon_0 \left(55 \epsilon_S^3+O\left(\epsilon_S^4\right)\right)+\epsilon_0^2 \left(\frac{2915 \epsilon_S^3}{54}+O\left(\epsilon_S^4\right)\right)+O\left(\epsilon_0^3\right)
\end{eqnarray}
By iterating this procedure many times, we numerically determined the region of state space that distills to the Norell state. This is shown in Figure \ref{triangle}. To translate this two-dimensional region into a single number, let us assume only depolarizing noise ($\epsilon_S=\epsilon_0=\delta_N/3$) on the input qutrits. We find the maximum depolarizing noise rate $\delta_N$ for input states to eventually distill to $\ket{N}$ is $0.38612$. This approximately, but not exactly, equal to the threshold for $\ket{S}$ state distillation. This threshold is substantially better than the threshold $0.32989$ for Norell states using the distillation protocol of \cite{Howard} which has only a linear error-supression.

 The region of state space that distills to the strange state is also shown in Figure \ref{triangle}. We could have used this twirling scheme for distilling the strange state. However this does not offer any advantages over the simpler twirling scheme for strange states discussed earlier.

 The probability of successful decoding of Norell states for small $\epsilon_0$ and $\epsilon_S$ is 
 \begin{equation}
     \mathcal P =\frac{1}{1728}-\frac{11}{1728}\epsilon_0 - \frac{11}{1728}\epsilon_S + \ldots.
 \end{equation} This results in a similar yield parameter, $\frac{1}{\log_3 19008} \approx .112$, as for distillation of $\ket{S}$ states. 
 
\section{Discussion}
\label{discussion}
The 11-qutrit Golay code distills strange states with a threshold to depolarizing noise of $\delta_*=0.38715$. This is the highest threshold of any known qutrit magic state distillation routine. Moreover, we emphasize that this threshold is a worst-case threshold that applies to all forms of noise, not just depolarizing noise, thanks to the twirling scheme presented above. The best threshold to depolarizing noise for a qubit magic state distillation routine is $\delta_*=0.34535$, which arises for distillation of $\ket{T}$ states via the 5-qubit code \cite{MSD}. So the 11-qutrit Golay code defines the first qutrit distillation protocol that also has a better threshold than any qubit distillation protocol, although it may not be meaningful to compare noise thresholds between qudits of different dimensionalities. Qudit codes for sufficiently large odd-prime dimension  \cite{campbell2014enhanced} do have higher thresholds to depolarizing noise, but, in these cases, the depolarizing noise threshold does not, on its own, completely characterize the distillable region of state space. 

This noise threshold is only a little over half of the theoretical upper limit for the noise threshold $\delta_*=3/4$, set by the necessity of contextuality (or positivity of the discrete Wigner function). Do other codes exist with better thresholds? We tried a similar construction with other self-orthogonal maximal ternary codes of length 11 and 13 \cite{selfdual1, selfdual2}; but the ternary Golay code is the only code we could find that is suitable for magic state distillation. At present, the 11-qutrit Golay code is the only code known to be able to distill the strange state. 

Magic state distillation with the 23-qubit Golay code was discussed briefly in \cite{Reichardt2005}, where it was shown that it is not suitable for distilling qubit $\ket{H}$ states. It is interesting to note that the 23-qubit Golay code is able to distill qubit $\ket{T}$ states, but with a threshold that is just slightly less than that of the 5-qubit code. As we review in appendix C, the error-suppression for $\ket{T}$ state distillation using the 23-qubit code is quadratic, as one would expect for a generic code. On the other hand, the ternary Golay code is the best known code for distillation of strange and Norell magic states, and is able to distill the strange state with a somewhat miraculous cubic error-suppression, whose origin needs to be better understood.


One motivation for distilling strange states is to address whether contextuality can be shown to be a sufficient resource for universal quantum computation.\cite{nature} This requires us to construct a distillation scheme that is tight to the boundary of the Wigner polytope, i.e., has a threshold to depolarizing noise of $3/4$. While it can be shown that no magic state distillation routine based on a finite stabilizer code can achieve this threshold, \cite{qudit-bound-states, bound}, the possibility remains that a sequence of stabilizer codes exist which distill the strange state, whose threshold  approaches $3/4$. Of course, the ternary Golay code is an extremely special error-correcting code, and there is no reason to expect that one can generalize it to obtain such a sequence of codes. Nevertheless, demonstrating the existence of a single magic state distillation routine that distills the strange state, is an important first step for this program.

\ack{
The author thanks Prof. Prem Saran Satsangi for inspiration and guidance. The author also thanks Mark Howard for comments on an earlier draft of the manuscript, and anonymous referees for valuable suggestions.} 

\funding{
This work is supported in part by a DST INSPIRE Faculty Award, DST-SERB Early Career Research Award (ECR/2017/001023) and MATRICS grant (MTR/2018/001077).}

\section*{Appendix A: State injection with the strange state}
\label{state-injection}
Let us show how the strange state can be used to implement a non-Clifford gate via state injection. We closely follow \cite{ACB}, where it was shown how the $\ket{N}$ and $\ket{H_{\pm 1}}$ can be used for state injection.

Let $U$ be a unitary operator whose eigenbasis is a complete set of stabilizer states. By a Clifford transformation, such an operator can be brought into a form where it is diagonal in the computational basis, 
\begin{equation}
    \begin{pmatrix}
     1 & 0 & 0 \\
     0 & e^{i\theta_1} & 0 \\
     0 & 0 & e^{i\theta_2}
    \end{pmatrix}.
\end{equation} We denote such an operator as $U_Z$. Such operators were referred to as ``equatorial operators'' in \cite{CampbellAnwarBrowne}. Define $\ket{U_Z}=U_Z H \ket{0}$,
\begin{equation}
\ket{U_Z(\theta_1,\theta_2)}=\frac{1}{\sqrt{3}} \left(\ket{0}+e^{i\theta_1}\ket{1}+e^{i\theta_2}\ket{2} \right). \label{diagonal-state}
\end{equation}
We refer to any state that can be brought into the above form via a Clifford unitary as an \textit{equatorial} state following \cite{ACB, CampbellAnwarBrowne}. $U_Z$ can be implemented by  state-injection circuit using  $\ket{U_Z}$ as follows: 
\begin{enumerate}
    \item Let qudit 1 be in the state $\ket{U_Z}$, and qudit 2 be in the state $\ket{\psi}$.
    \item Apply a controlled-$X^2$ gate to the $\ket{U_Z}\ket{\psi}$, with $\ket{\psi}$ as the target. 
    \item Measure $Z$ on qudit $2$; if the outcome is $\omega^m$, apply $(U_Z X^m U_Z^\dagger)$ to qudit 1. Qudit 1 is now in the state $U_Z \ket{\psi}$.
\end{enumerate}

The above procedure works if $U_Z^\dagger X U_Z$ is a Clifford operator, i.e., if $U_Z$ is in the third level of the Clifford hierarchy. \cite{Gottesman1999DemonstratingTV,DiagonalCliffordHierarchy} If $U_Z$ is not in the third-level of the Clifford hierarchy, then it is not possible to apply the outcome-dependent correction at the end. We then end up with the state $U_ZX^{-m} \ket{\psi}$ with a random, but known value of $m$. In this case, it is convenient to modify the circuit by applying another controlled-$X$, with qudit 1 as target, to obtain $X^m U_Z X^{-m} \ket{\psi}$. If $m=0$, we have obtained the desired state $U_Z\ket{\psi}$. If $m\neq 0$ we can repeat this state injection procedure in hopes of eventually reaching the state $U_Z\ket{\psi}$, or a state Clifford-equivalent to it. If the group $G$ generated by operators of the form $X^mU_ZX^{-m}$ is a finite group of relatively small (i.e., $\mathcal O(1)$) order, this process is a random walk which reaches $\ket{U_Z}$ in $\mathcal O(1)$ steps.

The magic states $\ket{N}$ and $\ket{S}$ are not equatorial states, but can be converted into equatorial states via a series of $2$-to-$1$ stabilizer reductions. \cite{ACB} showed how to convert the state $\ket{N}$ to an equatorial state via a $2$-to-$1$ stabilizer reduction: Start with two (very pure) qubits in the state $\ket{N}\ket{N}$. Project onto the codespace of the $[2,1]$ code defined by the stabilizer $\omega X_1 X_2$. This has a $1/4$ success probability. Decode treating $X_2$ as the logical $X$ operator, and $Z_1^2 Z_2$ as the logical $Z$ operator.

The resulting state is $X^2\ket{U_Z(\pi/3,2\pi/3)}$, which is Clifford equivalent to $\ket{U_Z(0,\pi)}$. $U_Z(0,\pi)$ is a non-Clifford gate; while it is not an element of the third-level of the Clifford hierarchy, the group generated by $X^{m}U_Z(0,\pi)X^{-m}$ is finite, and can be used to implement a non-Clifford gate as discussed above.

There is no $2$-to-$1$ stabilizer reduction which converts $\ket{S}$ to an equatorial state. However, we can convert two copies of an $\ket{S}$ state to a $\ket{N}$ state via the $2$-to-$1$ stabilizer reduction with stabilizer $Z_1 Z_2$, and decoding via logical operators $\bar{Z}=Z_2$ and $\bar{X}=X_1^2 X_2$. This stabilizer reduction succeeds with probability $1/2$. This stabilizer reduction can also convert two copies of any state of the form $\alpha \ket{1}+ \beta \ket{2}$, which is also a Clifford eigenstate \cite{jain}, into the state $\ket{N}$, with success probability $2 |\alpha \beta|^2$.)

This scheme appears to be the optimal scheme using only 2-qutrit stabilizer projections and Clifford unitaries. It would be interesting to search for schemes involving $n$-qutrit stabilizer projections, for $n>2$, that make better use of magic.

\section*{Appendix B: Distilled Norell states}

The output qutrit is in the state $\rho(\epsilon_0',\epsilon_S')$, where
\begin{equation}
    \epsilon_0'(\epsilon_0,\epsilon_S) = \epsilon_0^2 \frac{P_0(\epsilon_0,\epsilon_S)}{Q_N(\epsilon_0,\epsilon_S)}, ~~~ \epsilon_S'(\epsilon_0,\epsilon_S) = 6 e_S \frac{P_S(\epsilon_0,\epsilon_S)}{Q_N(\epsilon_0,\epsilon_S)},
\end{equation}
where
\begin{eqnarray}
P_0 & = & 55 - 495 \epsilon_0 + 1980 \epsilon_0^2 - 4092 \epsilon_0^3 + 3762 \epsilon_0^4 + 990 \epsilon_0^5 - 
  5940 \epsilon_0^6 + 5940 \epsilon_0^7  - 2673 \epsilon_0^8 \nonumber \\ && + 601 \epsilon_0^9 - 495 \epsilon_S + 
  3960 \epsilon_0 \epsilon_S - 13860 \epsilon_0^2 \epsilon_S + 24552 \epsilon_0^3 \epsilon_S - 18810 \epsilon_0^4 \epsilon_S - 
  3960 \epsilon_0^5 \epsilon_S \nonumber \\ && + 17820 \epsilon_0^6 \epsilon_S - 11880 \epsilon_0^7 \epsilon_S + 2673 \epsilon_0^8 \epsilon_S + 
  3960 \epsilon_S^2 - 27720 \epsilon_0 \epsilon_S^2 + 83160 \epsilon_0^2 \epsilon_S^2 
  \nonumber \\ && 
  - 122760 \epsilon_0^3 \epsilon_S^2  
  + 
  75240 \epsilon_0^4 \epsilon_S^2 + 11880 \epsilon_0^5 \epsilon_S^2 - 35640 \epsilon_0^6 \epsilon_S^2 + 
  11880 \epsilon_0^7 \epsilon_S^2 - 18480 \epsilon_S^3 \nonumber \\ && + 110880 \epsilon_0 \epsilon_S^3 - 277200 \epsilon_0^2 \epsilon_S^3  +
  327360 \epsilon_0^3 \epsilon_S^3 - 150480 \epsilon_0^4 \epsilon_S^3 - 15840 \epsilon_0^5 \epsilon_S^3 + 
  23760 \epsilon_0^6 \epsilon_S^3 \nonumber \\ && + 55440 \epsilon_S^4 - 277200 \epsilon_0 \epsilon_S^4 + 554400 \epsilon_0^2 \epsilon_S^4  -
  491040 \epsilon_0^3 \epsilon_S^4 + 150480 \epsilon_0^4 \epsilon_S^4 + 7920 \epsilon_0^5 \epsilon_S^4 \nonumber \\ && - 
  110880 \epsilon_S^5 + 443520 \epsilon_0 \epsilon_S^5 - 665280 \epsilon_0^2 \epsilon_S^5   + 
  392832 \epsilon_0^3 \epsilon_S^5 - 60192 \epsilon_0^4 \epsilon_S^5 + 147840 \epsilon_S^6 \nonumber \\ && 
  - 443520 \epsilon_0 \epsilon_S^6 + 
  443520 \epsilon_0^2 \epsilon_S^6 - 130944 \epsilon_0^3 \epsilon_S^6 - 126720 \epsilon_S^7  + 
  253440 \epsilon_0 \epsilon_S^7 - 126720 \epsilon_0^2 \epsilon_S^7 \nonumber \\ && + 63360 \epsilon_S^8 - 63360 \epsilon_0 \epsilon_S^8 - 
  14080 \epsilon_S^9 
\end{eqnarray}

\begin{dmath}
P_S  =   220 \epsilon_0^{10}+220 \epsilon_0^9 \epsilon_S-220 \epsilon_0^9-2585 \epsilon_0^8 \epsilon_S^2+3960 \epsilon_0^8 \epsilon_S-1980 \epsilon_0^8-11440 \epsilon_0^7 \epsilon_S^3+23320 \epsilon_0^7 \epsilon_S^2  
-17820 \epsilon_0^7 \epsilon_S+5940 \epsilon_0^7-21604 \epsilon_0^6 \epsilon_S^4+55000 \epsilon_0^6 \epsilon_S^3-56540 \epsilon_0^6 \epsilon_S^2+29040 \epsilon_0^6 \epsilon_S-7260 \epsilon_0^6 
-21032 \epsilon_0^5 \epsilon_S^5+64152 \epsilon_0^5 \epsilon_S^4-83160 \epsilon_0^5 \epsilon_S^3+58520 \epsilon_0^5 \epsilon_S^2-23100 \epsilon_0^5 \epsilon_S+4620 \epsilon_0^5-8800 \epsilon_0^4 \epsilon_S^6 
+31240 \epsilon_0^4 \epsilon_S^5-49500 \epsilon_0^4 \epsilon_S^4+46200 \epsilon_0^4 \epsilon_S^3-26950 \epsilon_0^4 \epsilon_S^2+9240 \epsilon_0^4 \epsilon_S-1540 \epsilon_0^4+1760 \epsilon_0^3 \epsilon_S^7
-7040 \epsilon_0^3 \epsilon_S^6+11440 \epsilon_0^3 \epsilon_S^5-7920 \epsilon_0^3 \epsilon_S^4+3080 \epsilon_0^3 \epsilon_S^2-1540 \epsilon_0^3 \epsilon_S+220 \epsilon_0^3+3520 \epsilon_0^2 \epsilon_S^8  -15840 \epsilon_0^2 \epsilon_S^7+31680 \epsilon_0^2 \epsilon_S^6-36080 \epsilon_0^2 \epsilon_S^5+24420 \epsilon_0^2 \epsilon_S^4-9240 \epsilon_0^2 \epsilon_S^3+1540 \epsilon_0^2 \epsilon_S^2+1408 \epsilon_0 \epsilon_S^9 
-7040 \epsilon_0 \epsilon_S^8+15840 \epsilon_0 \epsilon_S^7-21120 \epsilon_0 \epsilon_S^6+18040 \epsilon_0 \epsilon_S^5-9768 \epsilon_0 \epsilon_S^4+3080 \epsilon_0 \epsilon_S^3-440 \epsilon_0 \epsilon_S^2+256 \epsilon_S^{10}-1408 \epsilon_S^9+3520 \epsilon_S^8-5280 \epsilon_S^7+5280 \epsilon_S^6-3608 \epsilon_S^5+1628 \epsilon_S^4-440 \epsilon_S^3+55 \epsilon_S^2 
\end{dmath}

\begin{dmath}
Q_N  =  220 \epsilon_0^{11}+2475 \epsilon_0^{10} \epsilon_S-1155 \epsilon_0^{10}+10890 \epsilon_0^9 \epsilon_S^2-9900 \epsilon_0^9 \epsilon_S+4290 \epsilon_0^9+21120 \epsilon_0^8 \epsilon_S^3-26730 \epsilon_0^8 \epsilon_S^2+8910 \epsilon_0^8 \epsilon_S-6930 \epsilon_0^8+3960 \epsilon_0^7 \epsilon_S^4+5280 \epsilon_0^7 \epsilon_S^3-23760 \epsilon_0^7 \epsilon_S^2+19800 \epsilon_0^7 \epsilon_S+3960 \epsilon_0^7-63360 \epsilon_0^6 \epsilon_S^5+178200 \epsilon_0^6 \epsilon_S^4-224400 \epsilon_0^6 \epsilon_S^3+158400 \epsilon_0^6 \epsilon_S^2-60390 \epsilon_0^6 \epsilon_S+3366 \epsilon_0^6-132000 \epsilon_0^5 \epsilon_S^6+411840 \epsilon_0^5 \epsilon_S^5-574200 \epsilon_0^5 \epsilon_S^4+475200 \epsilon_0^5 \epsilon_S^3-247500 \epsilon_0^5 \epsilon_S^2+74448 \epsilon_0^5 \epsilon_S-7788 \epsilon_0^5-126720 \epsilon_0^4 \epsilon_S^7+448800 \epsilon_0^4 \epsilon_S^6-712800 \epsilon_0^4 \epsilon_S^5+693000 \epsilon_0^4 \epsilon_S^4-462000 \epsilon_0^4 \epsilon_S^3+207900 \epsilon_0^4 \epsilon_S^2-55440 \epsilon_0^4 \epsilon_S+6600 \epsilon_0^4-63360 \epsilon_0^3 \epsilon_S^8+253440 \epsilon_0^3 \epsilon_S^7-454080 \epsilon_0^3 \epsilon_S^6+506880 \epsilon_0^3 \epsilon_S^5-415800 \epsilon_0^3 \epsilon_S^4+258720 \epsilon_0^3 \epsilon_S^3-110880 \epsilon_0^3 \epsilon_S^2+27720 \epsilon_0^3 \epsilon_S-3300 \epsilon_0^3-14080 \epsilon_0^2 \epsilon_S^9+63360 \epsilon_0^2 \epsilon_S^8-126720 \epsilon_0^2 \epsilon_S^7+158400 \epsilon_0^2 \epsilon_S^6-158400 \epsilon_0^2 \epsilon_S^5+138600 \epsilon_0^2 \epsilon_S^4-92400 \epsilon_0^2 \epsilon_S^3+39600 \epsilon_0^2 \epsilon_S^2-9405 \epsilon_0^2 \epsilon_S+1045 \epsilon_0^2-5280 \epsilon_0 \epsilon_S^6+19008 \epsilon_0 \epsilon_S^5-27720 \epsilon_0 \epsilon_S^4+21120 \epsilon_0 \epsilon_S^3-8910 \epsilon_0 \epsilon_S^2+1980 \epsilon_0 \epsilon_S-198 \epsilon_0+1056 \epsilon_S^6-3168 \epsilon_S^5+3960 \epsilon_S^4-2640 \epsilon_S^3+990 \epsilon_S^2-198 \epsilon_S+18
\end{dmath}

Numerical basins computed in Figure \ref{triangle} appear to be symmetric with respect to interchange of the $\ket{N}$ and $\ket{S}$ state. This is not quite the case, as the thresholds to depolarizing noise for $\ket{S}$ and $\ket{N}$ states are slightly different. Interchange of $\ket{N}$ and $\ket{S}$ corresponds to interchange of $1-\epsilon_0-\epsilon_S$ and $\epsilon_S$. The expressions above are not symmetric under this exchange. 

\section*{Appendix C: Distillation with the 23-qubit Golay code}
Distillation with the 23-qubit Golay code was first reported in \cite{Reichardt2005}. There, it was found that 23-qubit Golay code is not suitable for distilling qubit $\ket{H}$ magic states, but it can distill $\ket{T}$ states. Here we present the results for  $\ket{T}$ state distillation in some more detail.

The 23-qubit Golay code \cite{PhysRevA.68.042322} is defined as the code given by the symplectic matrix:
\begin{equation}
\begin{pmatrix}
\mathbf{M}^{(2)}_c &|& 0 \\ 0 &|& \mathbf{M}^{(2)}_c \end{pmatrix}.
\end{equation}
where $\mathbf{M}_c^{(2)}$ is the binary generator matrix for the classical Golay code, as given in, e.g., \cite{PaetznickGolay}.

\begin{figure}
\centering
\includegraphics[width=.6\textwidth]{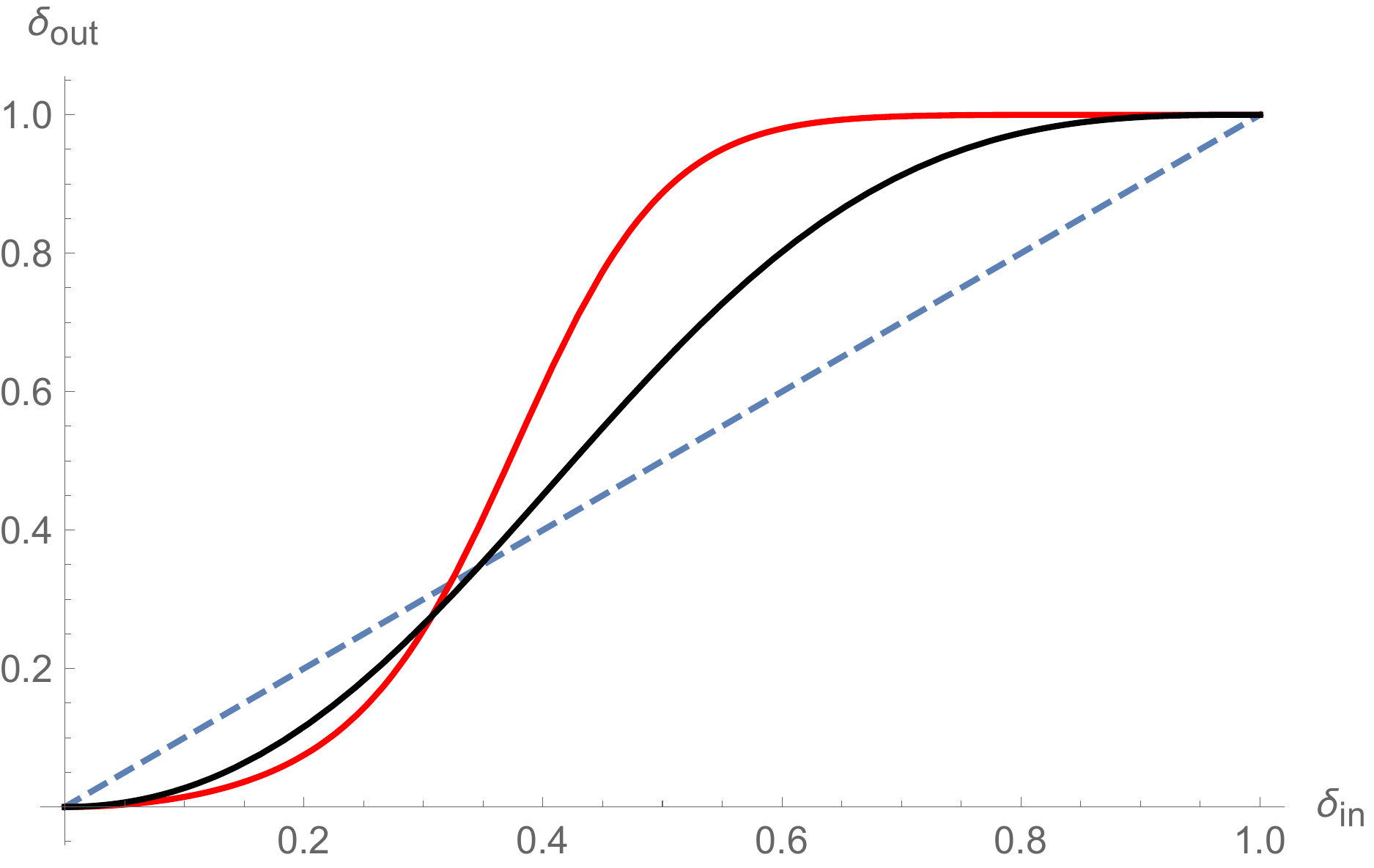}
\caption{The relation $\delta_{\text{out}}(\delta_{\text{in}})$ induced by distillation with the 23-qubit Golay code is shown by the solid red line. The dashed blue line is the line $\delta_{\text{out}}=\delta_{\text{in}}$, which is also shown for convenience, and the black line is the relation $\delta_{\text{out}}(\delta_{\text{in}})$ for the 5-qubit code. Both codes have quadratic reduction in noise, but the 5-qubit code has a better threshold.} \label{23-qubit-golay}
\end{figure}

The $\ket{T}$ magic state is defined as $\ket{T}\bra{T} = \frac{1}{2}(\mathbf{1} + \frac{1}{\sqrt{3}}(X+Y+Z))$, and is an eigenstate of the Clifford operator $T$ defined in \cite{MSD}. Noisy $\ket{T}$ states can be twirled to take the form:
\begin{equation}
\rho(\delta) = (1-\delta)\ket{T}\bra{T} + \delta\frac{\mathbf 1}{2}.
\end{equation}
Distilling with respect to the 23-qubit Golay code gives a relation $\delta_{\text{out}}(\delta)$ that takes the following form:
\begin{equation}
\delta_{\text{out}}= \delta^2 \frac{P_T(\delta)}{Q_{T}(\delta)} \approx \frac{253}{196}\delta^2
\end{equation}
where
\begin{dmath}
P_T(\delta) = 3895 \delta ^{21}-117921 \delta ^{20}+1297131 \delta ^{19}-6154225 \delta ^{18}+1514205 \delta ^{17}+142287453 \delta ^{16}-869243991 \delta ^{15}+2817045198 \delta ^{14}-5579251128 \delta ^{13}+5943010480 \delta ^{12}+978697104 \delta ^{11}-15862508256 \delta ^{10}+30813957440 \delta ^9-35023976064 \delta ^8+26000789760 \delta ^7-11870031360 \delta ^6+1942262784 \delta ^5+1403652096 \delta ^4-1189658624 \delta ^3+435240960 \delta ^2-87048192 \delta +8290304 
\end{dmath}
and
\begin{dmath}
Q_T(\delta) = -28336 \delta ^{22}+623392 \delta ^{21}-5801796 \delta ^{20}+28761040 \delta ^{19}-70542472 \delta ^{18}-19126800 \delta ^{17}+798925677 \delta ^{16}-3140863440 \delta ^{15}+7113803400 \delta ^{14}-10619737744 \delta ^{13}+10395332080 \delta ^{12}-5839214976 \delta ^{11}+931120960 \delta ^{10}-346508800 \delta ^9+4146253056 \delta ^8-8139005952 \delta ^7+8906118144 \delta ^6-6659186688 \delta ^5+3627008000 \delta ^4-1450803200 \delta ^3+410370048 \delta ^2-73859072 \delta +6422528.
\end{dmath}
This is plotted in Figure \ref{23-qubit-golay}.

Note that error-suppression is quadratic, as expected for a generic code of length $n=6m-1$, that has $T^{\otimes{n}}$ as a transversal operator. The threshold is at $\delta_*=0.32237$. This is slightly worse than the threshold of the 5-qubit code which is at $0.34535$. 
\bibliographystyle{RS}
\bibliography{qudit}

\end{document}